\documentclass[pre,preprint,showpacs]{revtex4}
\usepackage[T1]{fontenc}
\usepackage{ae}
\usepackage{epsfig}
\usepackage{graphicx}
\usepackage{bm}
\usepackage{amssymb}
\usepackage{amsmath}
\usepackage{amsthm}

\begin{document}
\title{Three- and four-state rock-paper-scissors games with diffusion}
\author{Matti Peltom\"aki and Mikko Alava}
\affiliation{Department of Engineering Physics, Helsinki University
of Technology, P.O.~Box 1100, 02015 HUT, Espoo, Finland}

\begin{abstract}
Cyclic dominance of three species is a commonly occurring
interaction dynamics, often denoted the rock-paper-scissors (RPS)
game. Such type of interactions is known to promote species
coexistence. Here, we generalize recent results of Reichenbach et
al. (e.g. Nature {\bf 448}, 1046 (2007)) of a four-state variant of
RPS. We show that spiral formation takes place only without a
conservation law for the total density. Nevertheless, in general
fast diffusion can destroy species coexistence. We also generalize
the four-state model to slightly varying reaction rates.  This is
shown both analytically and numerically not to change pattern
formation, or the effective wave length of the spirals, and
therefore does not alter the qualitative properties of the
cross-over to extinction.
\end{abstract}
\maketitle

\section{Introduction}

Pattern formation and stability of ecological multi-species systems
have attained a lot of interest recently \cite{aranson2002,
zhdanov2002, sole06,kerr2002,reichenbach2007}. These two topics are
usually linked together by the well-known fact that spatial
inhomogeneity can stabilize species co-existence
\cite{sole06,briggs04}, and understanding both the issues themselves
and their connection is of great importance.

An important class of model systems here are the so-called
rock-paper-scissors games. They describe the dynamics of three
species that cyclically dominate each other
\cite{szabo1999,szabo2004,szolnoki2004b,reichenbach2006,claussen2008}.
Typically, these systems involve a conservation law for the total
density. This can be imposed by writing rate equations such that the
densities of the three species always sum up to unity, or employing
lattice-based simulations where a site is always in exactly one of
the three states. In this case, the system has a reactive fixed
point which can be unstable, marginally stable, or stable depending
on other properties of the particular setting, such as discreteness
of the populations or dimensionality. There is a large body of work
on very similar models with different microscopic update rules
\cite{szabo2002,szolnoki2004,szolnoki2005,nishiuchi2008}, and
recently more complicated six-species systems with a similar
conservation law have been studied \cite{szabo2008}. Several
rock-paper-scissors--like processes have been identified in ecology,
both spatial \cite{kerr2002,keymer2006,loose2008} and non-spatial
\cite{sinervo1996,kirkup2004}, as well as in other contexts not
related to population dynamics such as the public goods game
\cite{semmann2003}.

Recently, Reichenbach and co-workers have studied a version of the
rock-paper-scissors game in which the conservation law has been
removed \cite{reichenbach2007,reichenbach2008}. In lattice
simulations, this is achieved by building the model on four states:
the three original cyclically dominating states and a fourth one
that denotes empty space. If diffusion is added,  it does not
interfere with the global conservation laws or absence thereof, but
serves to set a length scale different from that set by the lattice
constant. It has been shown that in this case the reactive fixed
point is always unstable, and with diffusion in two dimensions
spiral patterns form, similar to those in the complex
Ginzburg-Landau equation (CGLE) \cite{aranson2002}, or vortices in
ecological systems \cite{tainaka1989,tainaka1994}. The main
conclusion in Refs.~\cite{reichenbach2007,reichenbach2008} has been
that as a function of the diffusion constant $D$ there is a
cross-over from a reactive state with all three populations present
to an absorbing state in which only one of the populations survives.
This is argued to take place when the value of the diffusion
constant is such that the spirals outgrow the system size.

Here we analyze these models further by considering both the
four-state case and a three-state version with a conservation of the
total density. We show that removing the conservation law gives rise
to spiral formation that does not occur if the total density is
conserved. However, in spite of this, there is a mechanism involving
a diffusion-induced length scale that leads to a cross-over to an
absorbing state. Therefore, conclusions regarding population
stability are qualitatively the same for both cases. Second, all
previous theoretical studies have assumed that the microscopic
processes are rate-symmetric. In other words, one can cyclically
permute the three populations without any change whatsoever.
However, experiments can both explicitly show formation of spiral
patterns \cite{kerr2002} and still be built such that the detailed
pair-wise reaction mechanisms are qualitatively different for
different pairs of species \cite{loose2008}. It is apparent that a
small rate-asymmetry does not change these properties. We introduce
rate-asymmetry into the four-state model \cite{reichenbach2007}, and
argue analytically, that the resulting system is still essentially
the CGLE but only after an unimportant (in the first order) change
of variables. This result, also confirmed by direct simulations,
shows that already rate-symmetric theory and rate-asymmetric
experiments are comparable.

This paper is structured as follows. In Section II, we define both the
three- and the four-state models with and without diffusion, and discuss
in more detail what is previously known about them. In Section III, we
present our results for all cases considered. Finally, Section IV ends the
paper with a summary and discussion.

\section{The models and earlier work}

The three-state RPS model describes cyclic dominance of three states,
$A$, $B$, and $C$, and is
defined by the following reaction equations and corresponding rates
\begin{equation} \label{eq:reaction_three}
\begin{array}{ccccc}
AB & \rightarrow & AA & \mathrm{with} \; \mathrm{rate} &\mu \\
BC & \rightarrow & BB & \mathrm{with} \; \mathrm{rate} &\mu \\
CA & \rightarrow & CC & \mathrm{with} \; \mathrm{rate} &\mu \, .
\end{array}
\end{equation}
In the mean-field approximation, the rate equations describing the
evolution of the system are
\begin{equation} \label{eq:rate_eq_three}
\begin{array}{ccc}
\partial_t a & = & \mu a b - \mu a c \\
\partial_t b & = & \mu b c - \mu a b \\
\partial_t c & = & \mu c a - \mu c b \, ,
\end{array}
\end{equation}
where $a$, $b$, and $c$ are densities of the states $A$, $B$, and $C$, respectively. The rate
equations have a reactive fixed point at $a=b=c=\frac{1}{3}$, which is known to be
marginally stable. Such behavior is usually considered highly unrealistic in biologically
motivated systems, and therefore the MF approximation of this system is considered to be
of no relevance.

The same model with discrete densities (i.e.~finite populations)
with noise has been studied in the mean-field like, non-spatial
version \cite{reichenbach2006}. In this case, the marginal stability
of the fixed point remains, and the behavior of the system becomes a
random walk in the space of marginally stable circular trajectories.
Since the population is finite, this always leads to extinction in
the infinite-time limit.

However, in two-dimensional spatially extended systems the three-state RPS game is known to be
stable, at least in large enough systems. The simplest known theoretical approach reproducing the
behavior is the four-site approximation by Szabo and co-workers \cite{szabo2004}.

In empirical systems, the three states above are often considered to
be three cyclically dominating strains of bacteria. In such
settings, the dominance of strain $A$ over strain $B$ leads
microscopically to deaths of individuals of strain $B$. In these
cases, strain $A$ does not reproduce immediately but non-occupied
space is created, and the filling-in via reproduction of any strain
is a separate process. Therefore, the following set of reaction
equations has been proposed \cite{reichenbach2007}
\begin{equation} \label{eq:reaction_four}
\begin{array}{ccccc}
AB & \rightarrow & EA & \mathrm{with} \; \mathrm{rate} &\sigma \\
BC & \rightarrow & EB & \mathrm{with} \; \mathrm{rate} &\sigma \\
CA & \rightarrow & EC & \mathrm{with} \; \mathrm{rate} &\sigma \\
XE & \rightarrow & XX & \mathrm{with} \; \mathrm{rate} & \mu \, ,
\end{array}
\end{equation}
where $X$ can refer to any state and $E$ denotes empty space.
In contrast to Eqs.~(\ref{eq:reaction_three}) in which the only parameter $\mu$ sets the
time scale, these reaction equations contain two independent parameters. The corresponding
rate equations are \cite{reichenbach2008}
\begin{equation} \label{eq:rate_eq_four}
\begin{array}{ccc}
\partial_t a & = & a[\mu(1-\rho) - \sigma c]\\
\partial_t b & = & b[\mu(1-\rho) - \sigma a]\\
\partial_t c & = & c[\mu(1-\rho) - \sigma b]\, ,
\end{array}
\end{equation}
where $\rho=a+b+c$ is the total density. These equations have a reactive fixed point
\begin{equation} \label{eq:mf_fixed_point_four}
a=b=c=\frac{\mu}{3\mu + \sigma} \, ,
\end{equation}
which is linearly unstable for all $\mu$ and $\sigma$ \cite{reichenbach2008}.

To generalize both models to their spatially extended versions, let
the populations live on regular square lattices, the reactions take
place only in nearest-neighbour contact, and amend the reaction
equations with an exchange reaction
\begin{equation} \label{eq:exchange_reaction}
XY \rightarrow YX \; \; \mathrm{with} \;\; \mathrm{rate} \;\; D \, ,
\end{equation}
where $X$ and $Y$ can denote any state (including empty space in the four-state model),
and $D$ is the diffusion constant.

Exploiting the instability of the fixed point of Eq.~(\ref{eq:mf_fixed_point_four})
a spatially extended version of the system has been
recently approximatively mapped \cite{reichenbach2008}
to the complex Ginzburg-Landau equation (CGLE) \cite{aranson2002}.
In the mapping, one shifts the reactive fixed point to the origin,
expands around it to find a two-dimensional invariant manifold of the dynamics,
expresses the dynamics on the manifold, and performs a normal-form transformation,
i.e.~finds the quadratic transformation that is an identity mapping to first order
and the result of which has no quadratic terms. Upon adding the diffusion according
to Eq.~(\ref{eq:exchange_reaction}), one arrives at a variant of the CGLE.

In accordance with the known behavior of the CGLE, it has been found
that the spatial four-state model with diffusion leads to formation
of spirals \cite{reichenbach2007}. They have a characteristic wave
length that scales as the square root of the diffusion constant.
When the wave length is of the order of the system size, there is
essentially one spiral in the system. With even larger diffusion
constants, the system behaves essentially as a completely coupled
one and an extinction due to similar reasons as in the noisy
non-spatial case takes place \cite{reichenbach2006}. In other words,
there is an absorbing-state cross-over as a function of the
diffusion constant $D$ when the spiral wave length reaches the
system size.

In direct numerical simulations, all variants of the model are simulated on a regular
square lattice of size $L \times L$.
For each microscopic time step, a process (selection or diffusion in
the three-state case; selection, reproduction, or diffusion in the four-state case) is
chosen randomly with probabilities proportional to the rates, a random lattice site and
its random neighbour are chosen uniformly at random, and the reaction is executed if
allowed by the rules. The time is increased by $\Delta t = 1/(\tau L^2)$ where $\tau$ is the
sum of the rates for each case.
The procedure is repeated until the time $t$ reaches a
predefined value.

\section{Results}

\subsection{Three-state model}

\subsubsection{Mapping to the CGLE}

To answer the question if explicit handling of empty space,
i.e.~using the four-state model instead of the more traditional
three-state one, is really necessary to see the spiral pattern
formation and the consequent absorbing-state cross-over, we map the
three-state model to an equation resembling the CGLE as closely as
possible. To start, set the timescale in
Eqs.~(\ref{eq:rate_eq_three}) by choosing $\mu=1$. Introduce new
variables $x_A = a - \frac{1}{3}$, $x_B = b - \frac{1}{3}$, $x_C = c
- \frac{1}{3}$, and use the conservation of the total density
$a+b+c=1$ to express the dynamics in terms of $x_A$ and $x_B$ as
follows
\begin{eqnarray} \label{eq:rate_eqs_in_x}
\partial_t  x_A = \frac{1}{3} x_A + \frac{2}{3} x_B + x_A^2 + 2 x_A x_B \\
\partial_t  x_B = -\frac{2}{3} x_A - \frac{1}{3} x_B - 2 x_A x_B - x_B^2 \, .
\end{eqnarray}
Note that the treatment here differs from that of the four-state model, since there is
no need to reduce the number of dynamical variables by finding an invariant manifold.
From Eq.~(\ref{eq:rate_eqs_in_x}),
the linearization of the system around the fixed point is
\begin{equation}
\partial_t \left(\begin{array}{c} x_A\\x_B\end{array}\right) =
\left(\begin{array}{cc} \frac{1}{3} & \frac{2}{3} \\ -\frac{2}{3} & - \frac{1}{3} \end{array}\right)
\left(\begin{array}{c} x_A\\x_B\end{array}\right) \, .
\end{equation}
The eigenvalues of the linearization matrix are
\begin{equation}
\lambda = \pm i \frac{\sqrt{3}}{3} \, ,
\end{equation}
which recovers the known fact that the fixed point is marginally
stable.

To transform Eqs.~(\ref{eq:rate_eqs_in_x}) to a normal form, apply first the change of
variables
\begin{eqnarray} \label{eq:change_from_x_to_y}
y_A = \frac{2}{\sqrt{3}} x_A + \frac{1}{\sqrt{3}} x_B \\
y_B = x_B
\end{eqnarray}
to arrive at the rate equations, now expressed in terms of $y_A$ and $y_B$
\begin{eqnarray}\label{eq:rate_eqs_in_y}
\partial_t y_A = \frac{\sqrt{3}}{3}y_B + \frac{\sqrt{3}}{2}(y_A^2- y_B^2) \\
\partial_t y_B = -\frac{\sqrt{3}}{3}y_A - \sqrt{3} y_A y_B \, .
\end{eqnarray}
The reason for this particular change of variables will become clear when comparing the
normal form to be calculated below to the CGLE. Namely, this transformation ensures that
the linear terms are written in a directly comparable way.

Now, let us turn Eqs.~(\ref{eq:rate_eqs_in_y}) into their normal
form. Here, the task is to find such a quadratic transformation from
the variables $y_A$ and $y_B$ to a new pair of variables, say $z_A$
and $z_B$ such that the $y$'s and the $z$'s coincide to linear order
and that the rate equations for $z$ do not contain quadratic terms.
In general, finding such a transformation involves finding the
inverse of a particular $6 \times 6$ matrix, and particular corner
cases of singular matrices could cause the transform not to exist.
However, in the present example, this does not happen.

By imposing the restrictions above, we find the necessary transformation to be the following
\begin{eqnarray}
z_A = y_A + y_A y_B \\
z_B = y_B + \frac{1}{2} y_A^2 - \frac{1}{2} y_B^2
\end{eqnarray}
and the resulting normal form
\begin{eqnarray} \label{eq:normalform}
\partial_t z_A = \omega z_B + c_2 z_A^2 z_B + c_2 z_B^3 \\
\partial_t z_B = -\omega z_A  - c_2 z_A^3 - c_2 z_A z_B^2 \, ,
\end{eqnarray}
where $\omega = \frac{\sqrt{3}}{3}$ and $c_2 = -\frac{\sqrt{3}}{2}$ with fourth-order and
higher terms in $y_A$ and $y_B$ omitted.
To connect this to the
CGLE, consider the complex variable $z = z_A + i z_B$. Now,
Eqs.~(\ref{eq:normalform}) with added diffusion can be cast as
\begin{equation} \label{eq:cgleish}
\partial_t z = i \omega z + D \Delta z + i c_2 |z|^2 z \, .
\end{equation}
Further, changing to a rotating coordinate system (replace $z$ by
$z e^{i \omega t}$) removes the purely imaginary linear term, and the
resulting equation reads
\begin{equation} \label{eq:cgleish_nolin}
\partial_t z = D \Delta z + i c_2 |z|^2 z \, ,
\end{equation}
which is already close to the standard form of the CGLE
\begin{equation} \label{eq:standard_cgle}
\partial_t z = z + (1+ib)\Delta z - (1+ic)|z|^2 z \, .
\end{equation}
Comparing Eqs.~(\ref{eq:cgleish_nolin}) and (\ref{eq:standard_cgle})
reveals a crucial difference: in the CGLE there is a non-zero linear
term whereas in the normal-form complex PDE the three-state model
maps to, there is none. This is a direct consequence of the
stability properties of the original MF equations.

To further argue why omitting the linear term matters, consider a variant of the CGLE
with a more general linear term $\nu z$
\begin{equation} \label{eq:cgletilde}
\partial_{\tilde t} \tilde z = \nu \tilde z + (1+ib)\tilde \Delta \tilde z - (1+ic)|\tilde z|^2 \tilde z
\, .
\end{equation}
Given $\nu > 0$, Eq.~(\ref{eq:standard_cgle}) is recovered after the scalings
$\tilde z = \sqrt{\nu} z$, $\tilde t = 1/\nu t$, and $x = \sqrt{\nu} \tilde x$
\cite{aranson2002}. Substitute a generic single-spiral solution \cite{aranson2002}
\begin{equation} \label{eq:single-spiral-solution}
z = a(r) e^{i(\omega t \pm \phi + \psi(r))} \, ,
\end{equation}
where $a(r)$ and $\psi(r)$ are real functions and $r$ and $\phi$ are
the spatial coordinates in the cylindrical coordinate system, with
the spiral core at the origin, to Eq.~(\ref{eq:cgletilde}). Consider
the limit $r\to\infty$ assuming that $a(r)$ tends to a constant
(i.e.~the solutions are bounded) and that $\psi(r)$ tends to $qr$,
i.e.~asymptotically the spiral looks like a plane wave with wave
number $q$. One finds that for the single-spiral solution to exist,
the equality
\begin{equation} \label{eq:existence_condition}
a(r) \to a_0 = \sqrt{\nu-q^2}
\end{equation}
has to hold. Since $a(r)$ is real, spiral solutions with a finite wave length exist only for
positive $\nu$. In other words, there are no spiral solutions in the three-state RPS game. The same
argument can be repeated for plane waves as well, leading to exactly the same condition.

\begin{figure}[!h]
\begin{center}
\includegraphics[width=8cm]{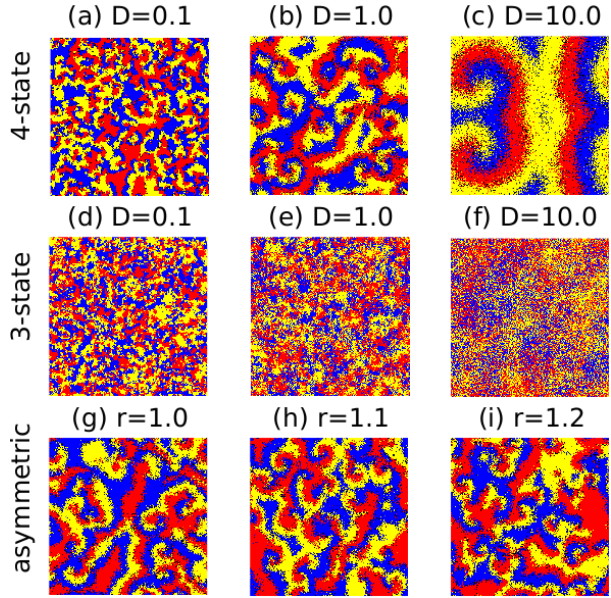}
\end{center}
\caption{(Color online)
Example runs of the system in a lattice of size $L \times L=200$.
(a), (b) and (c): the four-state
model with different diffusion constants and $\mu=\sigma=1$. (d), (e), (f): the three-state model
with different diffusion constants and $\mu=1$. (g), (h), (i): the rate-asymmetric four-state model
with different asymmetries $r = 1 + \epsilon$, $D=\mu=\sigma=1.0$.}
\label{fig:examplesystems}
\end{figure}

We have also verified the conclusion by direct numerical simulations
of the three-state model. Example configurations are plotted in
Fig.~\ref{fig:examplesystems} in panels (d), (e), and (f). There are
indeed no signs of spiral formation, but a growing diffusion
constant does change the spatial correlations in the system. Below,
we show that this leads to an absorbing-state cross-over as a
function of the diffusion constant regardless of the absence of
spiral pattern formation.

\subsubsection{Extinction cross-over}

\begin{figure}[!h]
\begin{center}
\includegraphics[width=8cm]{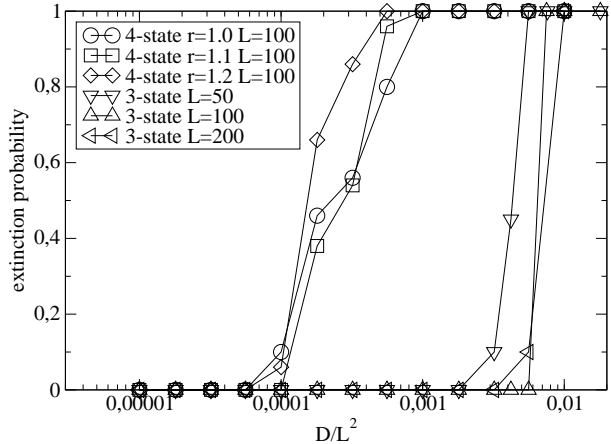}
\end{center}
\caption{The extinction probability as a function of the scaled
diffusion constant $D/L^2$ in various cases. For both the three- and
the four-state model there is a cross-over to an absorbing state
with two of the three subpopulations extinct. The location of this
cross-over scales as $D_c \propto L^2$ in both cases and in the
four-state case the introduction of rate-asymmetry does not affect
the location of the cross-over for small asymmetries.}
\label{fig:extinction_probability}
\end{figure}

Both the three-state and the four-state models have an emergent
length scale in the diffusive regime. In the four-state model, this
is the spiral wave length, and in the three-state case it is simply
the correlation length. Since the four-state model has an extinction
cross-over as a function of the diffusion constant because of the
length scale, it is natural to suspect that this is the case in the
three-state model as well. We have performed extensive numerical
simulations of both models to study the extinction probability as a
function of the diffusion constants. The results are shown in
Fig.~\ref{fig:extinction_probability}. For the four-state model we
recover the results of Ref.~\cite{reichenbach2007} and the results
for the three-state model show a similar cross-over. In both cases,
the diffusion constant at the cross-over scales as $D_c \sim L^2$
where $L$ is the linear dimension of the lattice.

\begin{figure}[!h]
\begin{center}
\includegraphics[width = 8cm]{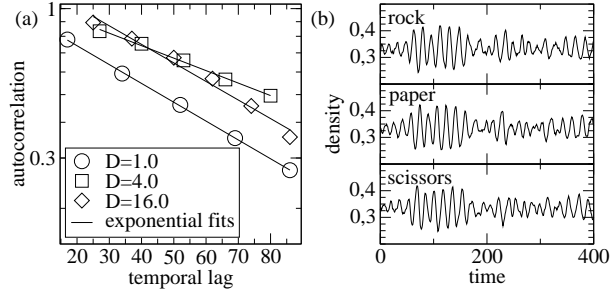}
\end{center}
\caption{(a) The envelope of the autocorrelation function of the density of state $A$
in the three-state model with different diffusion constants $D$. The symbols show the
numerical results and the solid black lines are fits to the form
$a(t) \propto e^{-t/t_d}$. For a wide range of diffusion constants we find
$t_d=66$ independent of $D$.
(b) The corresponding time series of the densities of the three states show oscillations
with wildly fluctuating amplitudes, often associated with almost-unstable dynamics and
time-scale separation \cite{peltomaki2008}.
The parameters are $L=200$ and $\mu=1.0$.}
\label{fig:autocorrelation}
\end{figure}

Next, we argue that the extinction cross-over is caused by the
presence of a correlation length, applying an argument previously
used in the context of the contact process with diffusion
\cite{marro1999,dantas2007,messer2008}. Imagine the system in its
steady-state, with a small spatially localized perturbation caused
by noise. Due to the exchange reactions, this perturbation diffuses
with diffusion constant $2D$, where the factor of two comes from the
fact that at each exchange reaction two particles moved. If the
typical lifetime of such a perturbation is $t_d$, it will diffuse up
to distance $x = \sqrt{2Dt_d}$ before decaying. After estimating
$t_d$, the cross-over should take place when $ L^2 = 2D t_d$ so that
the properly scaled diffusion constant at the cross-over becomes
\begin{equation}
\frac{D}{L^2} = \frac{1}{2t_d} \ .
\end{equation}
To estimate the perturbation lifetime, we have computed the autocorrelation function of the
time series of the densities. Fig.~\ref{fig:autocorrelation} shows the autocorrelation for
several diffusion constants $D$ and a corresponding time series. The immediate observations are
that the autocorrelation function is independent of $D$ and that it decays exponentially so
that a well-defined timescale exists. From the linear fits, we can extract the time scale and we
have found $t_d \approx 66$ (see Sec. II for the definition of unit of time in the simulations),
from which we get that at the cross-over we have
$\frac{D}{L^2} \approx 7.6 \cdot 10^{-3}$. This estimate corresponds well to the location of the
cross-over in Fig.~\ref{fig:extinction_probability}.

\subsection{Four-state model}

The previous formulations of both the three- and the four-state
models have assumed invariance of the model under cyclic
permutations of the three non-empty states. In particular, this is
seen in the rate equations (\ref{eq:rate_eq_three}) and
(\ref{eq:rate_eq_four}) where the reproduction rate $\mu$ is the
same for all species. However, this assumption is not necessarily
fulfilled by any realistic empirical scenario \cite{loose2008}. To
find out if this matters, we define a rate-asymmetric variant of the
four-state model by the following reaction equations
\begin{equation} \label{eq:reaction_four_asymmetric}
\begin{array}{ccccc}
AB & \rightarrow & AE & \mathrm{with} \; \mathrm{rate} &\sigma \\
BC & \rightarrow & BE & \mathrm{with} \; \mathrm{rate} &\sigma \\
CA & \rightarrow & CE & \mathrm{with} \; \mathrm{rate} &\sigma \\
AE & \rightarrow & AA & \mathrm{with} \; \mathrm{rate} & \mu \\
BE & \rightarrow & BB & \mathrm{with} \; \mathrm{rate} & \mu \\
CE & \rightarrow & CC & \mathrm{with} \; \mathrm{rate} & r\mu \\
XY & \rightarrow & YX & \mathrm{with} \; \mathrm{rate} & D \, ,
\end{array}
\end{equation}
where $r=1 + \epsilon$ ($\epsilon>0$) is assumed to be close to unity,
i.e.~$\epsilon \ll 1$,
and $X$ and
$Y$ can refer to any state (including the empty state $E$). This is the
simplest possible extension of the original four-state model that incorporates
rate-asymmetry. The corresponding MF rate equations are
\begin{equation} \label{eq:rate_eq_four_asymmetric}
\begin{array}{ccc}
\partial_t a & = & a[\mu(1-\rho) - \sigma c]\\
\partial_t b & = & b[\mu(1-\rho) - \sigma a]\\
\partial_t c & = & c[r\mu(1-\rho) - \sigma b] \, .
\end{array}
\end{equation}
These have the reactive fixed point
\begin{equation}
a = c = \frac{\mu}{(2+r)\mu+\sigma} \, ; \,
b = \frac{r\mu}{(2+r)\mu+\sigma} \, .
\end{equation}

To simplify further calculations, we would like to introduce such a set of three
dynamical variables $\tilde a$, $\tilde b$, $\tilde c$ so that the reactive fixed point
satisfies $\tilde a=\tilde b=\tilde c$. To this end, make the transformation
$\tilde a=a$, $\tilde b = b/r$, $\tilde c=c$ to arrive at the rate equations
\begin{equation} \label{eq:rate_eq_four_asymmetric_tilde}
\begin{array}{ccc}
\partial_t \tilde a & = & \tilde a[\mu(1-\rho) - \sigma \tilde c]\\
\partial_t \tilde b & = & \tilde b[\mu(1-\rho) - \sigma \tilde a]\\
\partial_t \tilde c & = & r \tilde c[\mu(1-\rho) - \sigma \tilde b] \, ,
\end{array}
\end{equation}
where $\rho = \tilde a + r\tilde b + \tilde c$, and the reactive
fixed point is $\tilde a = \tilde b = \tilde c = \frac{\mu}{(2+r)\mu
+ \sigma}$. As the first step of mapping these equations to partial
differential equations comparable to the CGLE, linearize around the
fixed point to arrive at the linearization matrix
\begin{equation}
L = -\frac{\mu}{D} \left( \begin{array}{ccc}
\mu & \mu r & \mu + \sigma \\
\mu + \sigma & \mu r & \mu \\
\mu r & \mu r^2 + \sigma r & \mu r \end{array} \right) \, ,
\end{equation}
where $D=(2+r)\mu + \sigma$. Now, we can make some quick
observations of the consequences of $r > 1$. First, the effect of
$\tilde b$ via the reproduction terms is multiplied by $r$ which is
natural since at the steady-state the increased reproduction of $C$
has to be balanced by an increased density of $B$ which dominates
over $C$ and therefore regulates it. Second, all time derivatives of
$\tilde c$ are multiplied by $r$ which can be understood so that the
faster reproduction of $C$ alters the intrinsic timescale.

From here, one would go further by first calculating the effect of
small $\epsilon$ on the two-dimensional invariant manifold of the
system by expressing the normal vector of the manifold at the fixed
point as its value in the rate-symmetric case and a first order
$\epsilon$-correction to it. From there, one can continue by
expressing the dynamics on the $\epsilon$-corrected manifold with
two dynamical variables, finding the corresponding normal form
transformation and applying it, assuming small $\epsilon$ whenever
necessary. While this procedure is certainly possible by brute force
(which we have certified by carrying it out), the resulting
expressions tend to get rather heavy since eigensystems of $3 \times
3$-matrices are involved, for instance. They do not appear to be
useful for gaining physical intuition, and thus we resort to a
qualitative argument of what the result, i.e.~the CGLE-like PDE,
necessarily is.

If $r=1$ ($\epsilon=0$), the resulting normal-form PDE is \cite{reichenbach2008}
\begin{equation} \label{eq:cgle_with_rmf_params}
\partial_t z = c_1 z + D\Delta z - c_2(1+ic_3)|z|^2 z \, ,
\end{equation}
where $z=z_A+ i z_B$ and $z_A$ and $z_B$ are the two dynamical
variables of the system. Here, the linear and cubic terms result
from the normal-form transformation, whereas the Laplacian term
comes from adding diffusion on top of the MF treatment. Similarly,
upon carrying out the procedure outlined above, the
$\epsilon$-perturbed terms are the linear term and the cubic term.
However, only the perturbations in the cubic term are relevant. This
can be justified as follows. Given general linear terms with four
$\epsilon$-dependent coefficients for a fixed $\epsilon$, one can
apply a linear transformation that diagonalizes the linear part. If
the resulting eigenvalues are not real, they can be made such by
transforming to a rotating coordinate system by the transformation
$z \mapsto z e^{i \omega t}$, as we did above to arrive at
Eq.~(\ref{eq:cgleish_nolin}). Then, either one of the variables can
be scaled such that the diagonal terms are equal. By these tricks,
the form of the linear term can always be cast to be as in
Eq.~(\ref{eq:cgle_with_rmf_params}). The remaining coefficient may
well be $\epsilon$-dependent, but this plays no role as to whether
the resulting equation has the properties and symmetries of the
CGLE.

So, the most general leading order correction from rate-asymmetry on top of
Eq.~(\ref{eq:cgle_with_rmf_params}) has to be of the form
\begin{widetext}
\begin{equation} \label{eq:cgle_nonsymm_correction}
\partial_t z = ... + \epsilon (d_1 z_A^3 + d_2 z_A^2 z_B + d_3 z_A z_B^2 + d_4 z_B^3)
+i\epsilon(d_5 z_A^3 + d_6 z_A^2 z_B + d_7 z_A z_B^2 + d_8 z_B^3) \, ,
\end{equation}
\end{widetext}
where the ellipsis stands for the non-perturbed terms, and the $d_i$'s are coefficients that
in the most general setting are functions of the parameters $\mu$ and $\sigma$ of the
four-state model. Now, to study the effect of the correction on spiral formation, let
us express the CGLE and its corrected counterpart in polar coordinates.
Write the phase-space coordinates as
\begin{equation}
z = R e^{i\theta}
\end{equation}
and the position-space coordinates as  $(r,\phi)$. The CGLE is now \cite{aranson2002}
\begin{equation}
\partial_t R = c_1 R + D(\Delta R - R(\nabla \theta)^2) - c_2 R^3
\end{equation}
\begin{equation}
R \partial_t \theta = D(2 \nabla\theta\cdot\nabla R + R \Delta \theta) - c_2 c_3 R^3 \, .
\end{equation}
And upon adding the nonsymmetric perturbation, we arrive at
\begin{equation} \label{eq:cgle_polar_radial}
\partial_t R = c_1 R + D(\Delta R - R(\nabla \theta)^2) - c_2 R^3 + \epsilon R^3 f(\theta)
\end{equation}
\begin{equation}
R \partial_t \theta = D(2 \nabla\theta\cdot\nabla R + R \Delta \theta) - c_2 c_3 R^3
+ \epsilon R^3 g(\theta) \, ,
\end{equation}
where $f$ and $g$ are smooth $2\pi$-periodic functions with the property that their Fourier
series do not have a constant term and have only a finite number of higher-order terms.

The single-spiral solution of the non-perturbed CGLE is given by
Eq.~(\ref{eq:single-spiral-solution}). To arrive at a similar
solution for the nonsymmetric case, consider first the corresponding
non-spatial dynamical system,
i.e.~Eqs.~(\ref{eq:cgle_with_rmf_params}) and
(\ref{eq:cgle_nonsymm_correction}) with the spatial derivatives
omitted. In this setting, both the symmetric and non-symmetric cases
produce limit cycles. In the symmetric case, they are circles, and
in the nonsymmetric case with small $\epsilon$ one can envision them
as perturbed circles. Then, the solution that is correct up to first
order in $\epsilon$ can be found by the change of variables that
transforms the perturbed circles back to regular circles. Such
change of variables in the first order in $\epsilon$ is of the
following form
\begin{equation}
R = \tilde{R}(1+\epsilon h(\theta))
\end{equation}
with the function $h(\theta)$ to be determined. Substituting this to
Eq.~(\ref{eq:cgle_polar_radial}) gives
\begin{displaymath}
0 = - F_0 q^2 [1+\epsilon h(\theta)]^2 + F_0(1-F_0^2) + \epsilon F_0^3 f(\theta)
\end{displaymath}
\begin{equation}
= -F_0 q^2 + F_0(1-F_0^2) +\epsilon(-2 F_0 q^2 h(\theta) + F_0^3 f(\theta)) + O(\epsilon^2) \, ,
\end{equation}
where the on the right-hand side (RHS) the $\epsilon$-independent part vanishes if $z(\tilde{r},\phi,t)$ is
a single-spiral solution of the CGLE, and the whole RHS vanishes up to first order
in $\epsilon$ if we choose
\begin{equation}
h(\theta) = \frac{F_0^2}{2q^2} f(\theta) \, .
\end{equation}

As a conclusion,
the perturbed system has a single-spiral solution where the oscillation amplitude
has an additional phase-dependent prefactor that essentially cancels out the
perturbation to the circular form of the limit cycle of the corresponding non-spatial
system. This does not change the wave length of the spirals (averaged over $\theta$)
which, in turn, means that the value of the diffusion constant at which the spirals
outgrow the system size does not change, so that one expects that a small
asymmetry does not change the location of the extinction crossover.

\begin{figure}[!h]
\begin{center}
\includegraphics[width = 4cm]{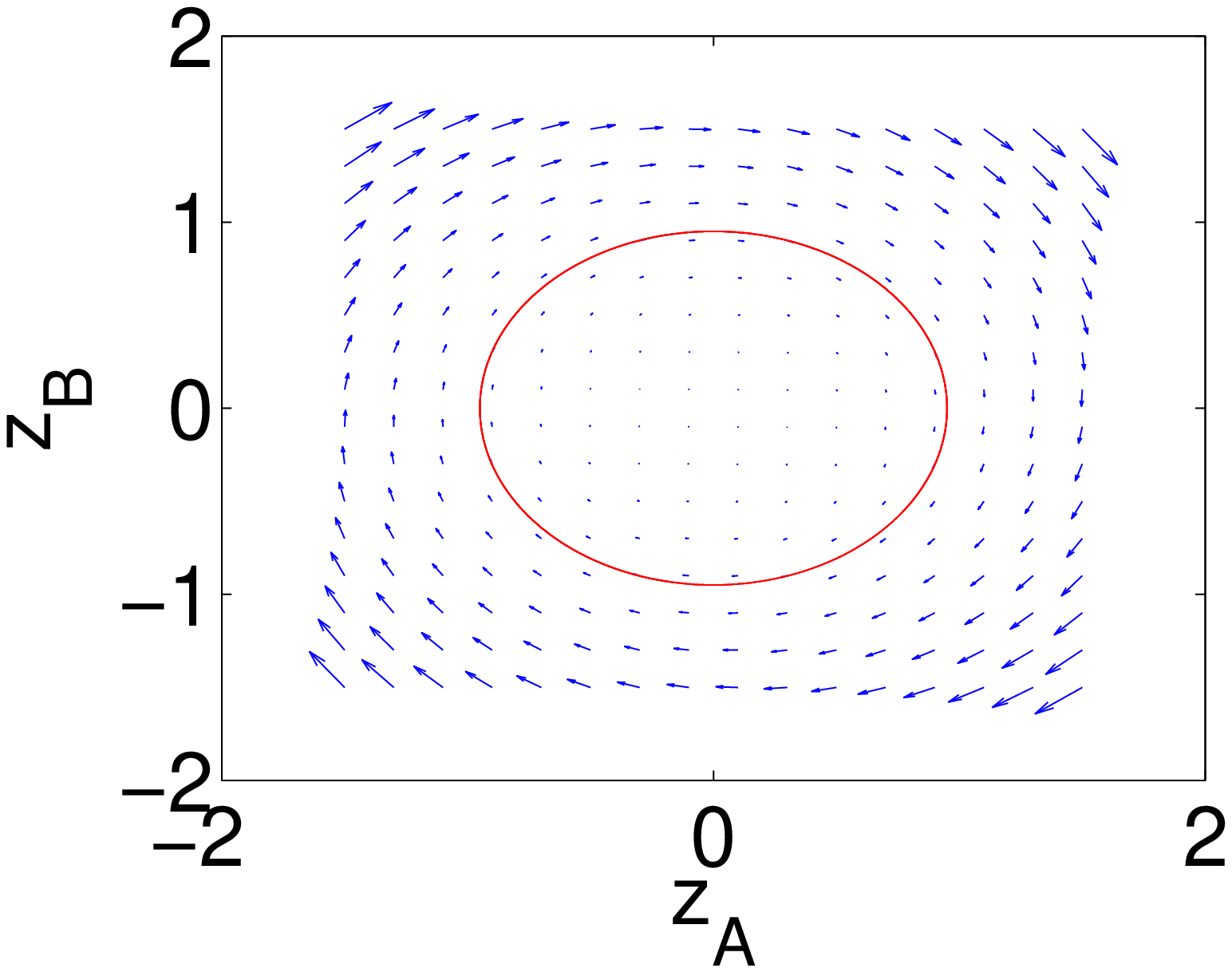}
\includegraphics[width = 4cm]{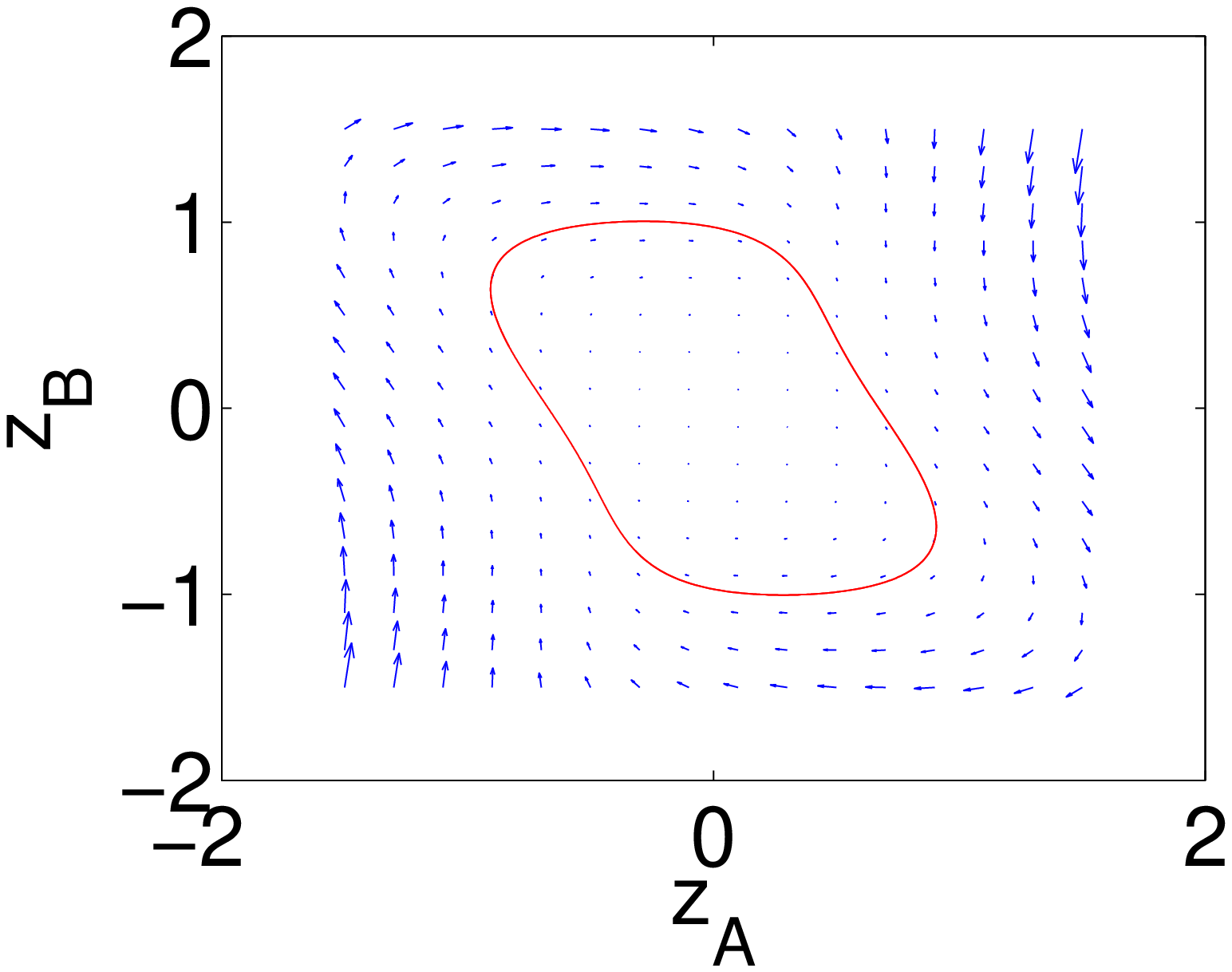}
\end{center}
\caption{(Color online)
The phase-space flow and limit cycle trajectories for the non-spatial rate-asymmetric
four-state model with $r=1.0$ (on the left) and $r=1.06$ (on the right).
The parameters are $\mu=\sigma=1$.}
\label{fig:trajectories}
\end{figure}

We have verified this prediction also numerically. First, panels (g), (h), and (i) of
Fig.~\ref{fig:examplesystems} show direct simulations of the four-state model with
increasing values of the asymmetry parameter $r$. One sees that the visual appearance
of the patterns is modified due to the phase-dependent prefactor $h(\theta)$ but that
the effective wave length appears to undergo no changes. Furthermore, we have
systematically varied the diffusion constant $D$ for different values of $r$ studying
the extinction probability. The results are shown in
Fig.~(\ref{fig:extinction_probability}). The
evident conclusion is that for small $\epsilon$
the cross-over
remains untouched. Finally, we have
visualized the perturbation to the limit cycles of the non-spatial dynamical system
in Fig.~(\ref{fig:trajectories}).

\section{Discussion}

In this work, we have addressed two questions related to variants of
the rock-paper-scissors game in which the total density is not
conserved. First, we have shown that the spiral formation observed
in the four-state case is a property of the four-state case only,
and does not generalize to the more widely studied three-state
rock-paper-scissors game with conserved total density. The spiral
formation has been previously argued \cite{reichenbach2007} to lead
to a cross-over from coexistence of all three populations to an
absorbing state with one surviving population and the other two
extinct. The mechanism has been considered to be explicitly
spiral-formation induced since the cross-over takes place at exactly
that value of the parameters where the spiral wave length becomes
equal to the linear size of the system. Here, we have shown that a
similar cross-over takes place in the three-state model as well even
though there are no signs of spiral (or any other visible pattern)
formation at all. We have identified that the mechanism of the said
cross-over is the appearance of a length scale from the interplay
between fluctuations and diffusion.

This result has direct consequences regarding the stability of such systems or, put in other
words, the extent to which biodiversity is maintained. Namely, the message is that in cyclically
dominating systems of three species, fast diffusion or big mobility can destroy species
coexistence regardless of whether there is a conservation law of total density in the system or not.
The mechanism can either show up as visible pattern formation (here, the four-state case) or
not (the three-state case). However, attention has to be paid to the details of the system if
quantitative predictions are to be made. The value of the diffusion constant at which the
destruction of coexistence is observed can differ by more than an order of magnitude for
different cases (see Fig.~\ref{fig:extinction_probability}). The result also hints to the
direction that the law of conservation of total density might play a role in pattern
formation in more complicated cases as well (see Ref.~\cite{szabo2008}, for instance).

Furthermore, the previous studies have considered only the cases
where the reaction rates between all pairs of species are equal.
This could potentially be a serious limitation since such equalities
do not exist in reality, generally \cite{loose2008}. We have
extended the previously defined four-state model
\cite{reichenbach2007} to cases where the reaction rates are not
equal. By looking at the first-order perturbation from the case with
equal densities, we have been able to map this case to a partial
differential equation that, in turn, can be turned into the complex
Ginzburg-Landau equation with a transformation of variables that
does not alter the qualitative properties of the system in the first
order. In particular, the effective wave length of the spirals
equals that of the case with equal densities. As a conclusion, a
small asymmetry in the reaction rates does not change either the
properties of the pattern formation nor the location of the
cross-over to the absorbing state. Altogether, this result serves to
give a partial explanation why pattern formation and cross-over to
extinction are visible in experiments as well. The behavior at
larger asymmetries remains to be explored.

\section*{Acknowledgements}

This work has been supported by the Academy of Finland through the Center of Excellence program.

\end{document}